\documentstyle[12pt]{article}
\begin{document}
\pagestyle{plain}
\title{The impact force in the ballistic oscillator}
\author{Miroslav Pardy\\
Department of Physical Electronics \\
Masaryk University \\
Kotl\'{a}\v{r}sk\'{a} 2, 611 37 Brno, Czech Republic\\
e-mail:pamir@physics.muni.cz}
\date{\today}
\maketitle
\vspace{50mm}

\begin{abstract}
We determine the impact force generated by the projectile  penetrating the body of the harmonic oscillator.We discuss the ballistic effect in connection with the Compton effect. 

\end{abstract}
\vspace{1cm}

\section{Introduction}
The ballistic pendulum is well known in the military science where the velocity of bullet or projectile was measured by such pendulum.
Leonard Euler (Euler, 1745; Euler,1961) discusses the physical ballistic pendulum in his monograph on artillery.
In general, the projectile of mass $\mu$ is captured by the body of mass  $m$ of the pendulum and the system with mass $\mu + m$ oscillates with amplitude A. We show here how to determine the force which is result of the interaction of the projectile with the body of the pendulum. The knowledge of physical laws enables to determine the velocity of projectile from the mechanical motion  of the pendulum. To be pedagogical clear let us start from the elementary theory of mathematical and physical ballistic pendulum.    

\section{The mathematical ballistic pendulum}

The mathematical ballistic pendulum is realized by the mathematical pendulum  with mass $m$ and with the projectile  with mass $\mu$. The initial velocity of the projectile is supposed to be $v$. After the interaction of the projectile with the massive body of the pendulum, the final velocity of the system $m + \mu$ let be $v'$. The conservation of the momentum is written in the following relation:

$$(m + \mu)v' = \mu v .\eqno(1)$$

The potential energy of the system after collision is 

$$E_{p} = (m + \mu)gh, \eqno(2)$$
where $h$ is the vertical distance  of the system $\mu + m$ from the horizontal plane z =0.  The kinetic energy of the system after collision is 

$$E_{k} = \frac{(\mu + m)v'^{2}}{2} = \frac{\mu^{2} v^{2}}{2(\mu + m)},\eqno(3)$$
where we used equation (1). Equalization of the potential energy with the kinetic energy leads to the equation for the initial velocity of the projectile. Or, 

$$v = \frac{m + \mu}{\mu}\sqrt{2gh}.\eqno(4)$$
By the experimental determination of  $h$, we get the velocity of the projectile with mass $\mu$.

\section{The physical ballistic pendulum}

The physical ballistic pendulum is realized by the physical pendulum  with momentum of inertia $J_{0}$, mass $M$  and with the projectile  with mass $\mu$. The initial velocity of the projectile is supposed to be $v$. After the interaction of the projectile with the massive body of the pendulum, the final angular velocity of the system $M + \mu$ let be $\omega$. The conservation of the angular momentum is written in the following relation (with $r$ being the distance of the projectile to the axis or rotation).

$$(J_{0}+ \mu r^{2})\omega = \mu vr,\eqno(5)$$
from which follows

$$v = \frac{(J_{0}+ \mu r^{2})\omega}{\mu r}.\eqno(6)$$

The rotation energy of the pendulum is 

$$ E _{rot} = \frac{1}{2}(J_{0}+ \mu r^{2})\omega^{2}.\eqno(7)$$

If $\lambda$ is the distance 
of the center of inertia to the point of rotation, then the potential energy of the system after collision is 

$$E_{p} = Mg\lambda(1 -\cos\varphi) + \mu g r (1 - \cos\varphi). \eqno(8)$$

The equalization of the potential energy with the rotation energy kinetic energy (7) leads to the equation for the final angular velocity of pendulum. Or, with $1-\cos\varphi = 2\sin^{2}(\varphi/2)$,

$$\omega  = 2 \sin(\varphi/2)\sqrt{\frac{g(M\lambda + \mu r)}{J_{0} + \mu r^{2}}}.\eqno(9)$$

Then, after insertion of $\omega$ into the formula for velocity we
get 

$$v = \frac{2\sin(\varphi/2)}{\mu r} \sqrt{g(M\lambda + \mu r)(J_{0} + \mu r^{2})}.\eqno(10)$$

By the experimental determination of  an angle $\varphi$, we get the velocity of the projectile with mass $\mu$.

\section{The stimulated harmonic oscillator}

If the external force is $F(t)$, then the Newton equation for the motion of harmonic oscillator is 

$$\ddot x + \omega^2 x = \frac{1}{m}F(t); \quad\omega = \sqrt{\frac{k}{m}},\eqno(11)$$
where $k$ is the elastic constant following from the definition of the potential energy of harmonic oscillator.

It is suitable to solve the last equation as follows (Landau et al., 1965). We write the initial equation in the form

$$\frac{d}{dt}(\dot x +i\omega x) -i\omega(\dot x + i\omega x) = 
\frac{1}{m}F(t),\eqno(12)$$
or, 

$$\frac{d\xi}{dt} - i\omega\xi = \frac{1}{m}F(t),\eqno(13)$$
where we have introduced the new quantity

$$\xi = \dot x + i\omega x. \eqno(14)$$

We suppose the solution in the form

$$\xi(t) = A(t)e^{i\omega t}. \eqno(15)$$

After insertion of this form to the equation (13), we get 

$$\dot A(t) = \frac{1}{m}F(t)e^{-i\omega t}\eqno(16)$$
and the quanity $\xi$ is as follows:

$$\xi(t)  = e^{i\omega t}\left\{\int_{0}^t \frac{1}{m}F(t)e^{-i\omega t} dt \right\} + \xi_{0},\eqno(17)$$
where $\xi_{0}$ is the initial value of $\xi$, or, $\xi_{0} = \xi(0)$.

The real motion is expressed by the coordinate $x$ which follows from eq. (14) as

$$x(t) = \frac{1}{\omega}{\rm Im}\;\xi(t).\eqno(18)$$

\section{The short half-periodic pulse stimulation}

Let us consider the situation, where the oscillator is stimulated by the very short half-period sinus force. In other words, the equation of motion is as follows:

$$\ddot x +  \omega_{0}x = \frac{1}{m}F_{0}\sin\Omega t; \quad 0 <
t < \pi/\Omega; \quad \omega_{0} = \sqrt{k/m}. \eqno(19)$$

It follows from the preceding section that the solution is of the form

$$\xi(t)  = e^{i\omega_{0} t}\left\{\int_{0}^{\pi/\Omega} \frac{F_{0}}{m}\frac{1}{2i}
\left(e^{i\Omega t} - e^{-i\Omega t}\right)e^{-i\omega_{0} t} dt \right\} + \xi_{0}.\eqno(20$$

After some elementary integration we get the solution in the form
(with $\xi_{0} = 0$)
$$\xi(t) = - \frac{F_{0}}{2m}\frac{1}{\omega_{1}\omega_{2}}\quad \times $$

$$\left[\omega_{2}(\cos\lambda_{1} + i\sin\lambda_{1}) + 
\omega_{1}(\cos\lambda_{2} + i\sin\lambda_{2}) - (\omega_{2} + \omega_{1})(\cos\omega_{0}t + i\sin\omega_{0}t\right], \eqno(21)$$
from which follows the real and imaginary parts in the form:

$${\rm Re}\;\xi = - \frac{F_{0}}{2m}\frac{1}{\omega_{1}\omega_{2}}
\left[\omega_{2}\cos\lambda_{1} + 
\omega_{1}\cos\lambda_{2} - 2\Omega\cos\omega_{0}t \right], \eqno(22)$$

$${\rm Im}\; \xi =  -\frac{F_{0}}{2m}\frac{1}{\omega_{2}\omega_{2}}
\left[\omega_{2}\sin\lambda_{1}  + \omega_{1}\sin\lambda_{2} - 2\Omega\sin\omega_{0}t\right], \eqno(23)$$
where constants $\omega_{1},\omega_{2}$ and $\lambda_{1},\lambda_{2}$ were calculated as follows:

$$\omega_{1} = \Omega - \omega_{0},\quad \omega_{2} = \Omega + \omega_{0},  \eqno(24)$$

$$\lambda_{1} = \omega_{0}t + \frac{\omega_{1}}{\Omega}\pi, \quad \lambda_{2} = \omega_{0}t - \frac{\omega_{2}}{\Omega}\pi.  \eqno(25)$$

After absorption of the  projectile by the oscillator mass, the 
ballistic oscillator will oscillate with frequency 

$$ \omega = \sqrt{\frac{k}{m + \mu}}.\eqno(26)$$

The comparison of eq. (18) with the experimental motion of the oscillator, we can determine the parameters $F_{0}$ and $\Omega$ of the short pulse force impacting the oscillator.  
 
\section{The linear pulse stimulation}

Let us consider the stimulated force in the form 

$$F(t) = F_{0} - at; \quad 0 < t < F_{0/a} \eqno(27)$$
which is of more realistic  meaning because the  projectile  penetrating into some medium is continually bremsed to stop after  some time. In other words the initial force is zero after  time $F_{0}/a$. The explicit of equation (17) with force (27) is as follows:

$$\xi(t)  = e^{i\omega_{0} t}\left\{\int_{0}^{F_{0}/a} \frac{F_{0} - at}{m}e^{-i\omega_{0} t} dt \right\} + \xi_{0},\eqno(28)$$

After some elementary integration and algebraic modification, we get the following result:

$$\xi(t)  = -\frac{iF_{0}}{m\omega_{0}}e^{i\omega_{0}t} -\frac{a}{m\omega_{0}^{2}}\left(e^{i\lambda} - e^{i\omega_{0}t} \right),\eqno(29)$$
where 

$$\lambda = \omega_{0}t - \frac{\omega_{0}F_{0}}{a}.\eqno(30)$$
The motion of the oscillator is expressed by the coordinate $x$. Or,

$$ x = \frac{1}{\omega_{0}}{\rm Im}\; \xi =  \left(-\frac{1}{m\omega_{0}^{3}}\right)(a\sin\lambda  - a\sin \omega_{0}t + F_{0}\omega_{0}\cos\omega_{0}t);
\quad 0 < t < F_{0}/a.\eqno(31)$$

The motion after time $F_{0}/a$ is harmonic with frequency $\omega = \sqrt{k/(m + \mu)}$. The experimental motion can be compared with the theoretical one in order to get the parameters of stimulated force which is generated by the projectile interaction with the massive body of the pendulum.

\section{Discussion}

The article is a modification  of the classical problems  of ballistic vibrational systems such as it is the ballistic mathematical and physical pendulum. The modification consists in using so called ballistic harmonic oscillator. We do not consider the damped mathematical and physical pendulum, or the damped harmonic oscillators.

We have seen how to determine the force which is generated by some projectile  when it is penetrating through mass of harmonic oscillator. The problem of the ballistic pendulum is discussed by Euler in his famous book on the artillery (Euler,1745); Euler,1961) and it is not discussed in his monographs on classical mechanics (Euler, 1736; Euler, 1715). The  book on the Newton mechanics written by Tait and Steele (Tait et al. 1889)  does not solve the problem of the ballistic pendulum, or ballistic oscillator. So, our article is in some sense original one. The forensic detectives are interested  in the problems concerning ballistics. So, it is not excluded that the ideas in our article will be used by the forensic investigators.

The ideas of the article can be extended to quantum electrodynamics. One example can be the Compton effect described by the symbolic equation

$$\gamma  + e  \quad  \rightarrow  \quad  \gamma + e \eqno(31)$$

According the classical approach, presented in the preceding text the electron is accelerated by some force in order to get the final velocity. The force cannot be determined by the methods of quantum electrodynamics. In other words such force is Kantian "{\it Ding an sich}". At the same time, if electron is accelerated, then it must generate the electromagnetic energy according to the Larmor formula(Landau et al., 1988):

$$\frac{dE}{dt} =  \frac{2}{3}\frac{e^{2}}{c^{3}}\left(\left|\frac{d^{2}x}{dt^{2}}\right|
\right)^{2}\eqno(32)$$

However, the Compton formula is exact one and no additional correction in the form of the electromagnetic radiation is involved in it. It means that electron is not accelerated from  the viewpoint of the classical electrodynamics. This is not contradiction, because we know that the quantum world differs from the world of the common sense. 
 
 It was published many times that accelerated charge in vacuum detects so called Unruh temperature of vacuum. Electron is accelerated by photon during the Compton process but no Unruh temperature is generated. So the classical process is not in harmony with the quantum one. The Compton formula can be also calculated from the Volkov solution (Volkov, 1935) of the Dirac equation in quantum electrodynamics (Nikishov, 1979; Ritus, 1979; Berestetzkii, 1989; Pardy, 2003; Pardy, 2004). However, no Larmor process and Unruh process follows from this modern approach.  
 
The Compton process has an analogue in the elementary particle physics in LHC where the interaction of protons with the installed laser gun leads to the Compton effect with protons:

$$\gamma  + p  \quad  \rightarrow  \quad  \gamma + p \eqno(33)$$

In this case, proton is composed from three quarks and so we feel, a priori,  that the generation of the internal force initiated by photon in the proton is physically meaningful. The process (33) can be at the same time considered as the two-step process, where the first part is $\gamma  + p  \rightarrow  p^{*}$, where $p^{*}$ is the excited state of proton, and the second part is $p^{*}  \rightarrow  \gamma + p$. Such photo-nuclear reactions are studied in the more general form in many articles.

Very interesting analog of the equations (31, 33) is the interaction of photons with the fullerene $C_{60}$:

$$\gamma  + C_{60}  \quad  \rightarrow  \quad  \gamma + C_{60}, \eqno(34)$$
where fullerene is the gigantic molecule composed from sixty carbons. In this case it is possible to consider the individual process 
$\gamma  + C_{60} \rightarrow  C^{*}_{60}$, where $C^{*}_{60}$ is the excited state of the fullerene molecule, and the second step $C^{*}_{60} \rightarrow \gamma + C_{60}$. The Lebed\v ev light pressure theory can be also applied here because the fullerene is in some approximation the classical optical object.   
It is possible to consider also the photoelectric effect, which is still under discussion and the internal atomic Compton effect which represents very difficult problem of the quantum fullerene physics.

The interaction of photons with graphene is possible in the form of the Compton process

$$\gamma  + G  \quad  \rightarrow  \quad  \gamma + G \eqno(35)$$

The interaction of photon with graphene also involves the photoelectric effect at the low energies of photons.

All ballistic interactions can be considered in the magnetic field, in such a way forming many outstanding  problems waiting to be resolved.
 
\vspace{7mm}
 
\noindent
{\bf References}

\vspace{7mm}

\noindent
Berestetzkii V. B., Lifshitz E. M. and  Pitaevskii L. P.,
{\it Quantum Electrodynamics}, Moscow, Nauka, 1989. (in Russian). \\[2mm]
Euler, L. {\it Neue Grunds\"atze der Artillerie }, Berlin, K\"onigl. und der Academie der
Wissenschaften privil. Buchh\"andler, Berlin 1745.; ibid. Euler Opera Omnia, 1922., 2. Reihe, V.XIV. \\[2mm]
Euler, L. {\it New foundation of artillery}, Moscow, GIFML, 1961. (in Russian). \\[2mm]
Euler, L.  {\it Mechanica sive motus scientia analytice exposita auctore Leonhardo Eulero}, Tomus I., Petropoli, Ex Typographia Academiae Scientiarum, A. 1736. \\[2mm]
Euler, L. {\it Theoria motus corporum solidorum seu rigidorum}, Rostochii et Gryphiswaldiae, MDCCXV. \\[2mm]
Landau, L. D. and  Lifshitz, E. M. {\it Mechanics}, 2-nd ed., Moscow, Nauka, 1965. (in Russian). \\[2mm]
Landau, L. D. and  Lifshitz, E. M. {\it The classical theory of fields}, 7-th ed., Moscow, Nauka, 1988. (in Russian). \\[2mm]
Nikishov, A. I. (1979). The problem of the intensive external field in quantum electrodynamics, Trudy FIAN 111, 152. (in Russian).\\[2mm]
Pardy, M. (2003). Electron in the ultrashort laser pulse,
International Journal of Theoretical Physics, {\bf 42}(1), 99.\\[2mm]
Pardy, M. (2004). Massive photons and the Volkov solution, International Journal of Theoretical Physics, {\bf 43}(1) 127.\\[2mm]
Ritus, V. I. (1979). The quantum effects of the interaction of elementary particles with
the intense electromagnetic field, Trudy FIAN 111, pp. 5-151. (in Russian).\\[2mm]
Tait, P. G. and Steele, W. J. {\it Dynamics of a particle}, 6-th ed. Macmillan \& Company and New York, 1889. \\[2mm]
Volkov, D. M. (1935). $\ddot{\rm U}$ber
eine Klasse von L$\ddot{\rm o}$sungen
der Diracschen Gleichung,\\ Zeitschrift f$\ddot{\rm u}$r Physik,
{\bf 94}, 250.
 
\end{document}